\documentclass[twocolumn]{article}  
\usepackage{graphicx}
\usepackage[T1]{fontenc}
\usepackage[subrefformat=parens]{subcaption}

\makeatletter
\let\cl@chapter\undefined
\makeatletter

\usepackage{amsmath,amssymb}   
\usepackage{amsthm} 
\usepackage{mathtools} 
\usepackage[numbers,sort&compress]{natbib}
\usepackage[margin=20mm]{geometry}
\usepackage{hyperref}
\hypersetup{colorlinks=true,pdfborder={0 0 0}}
\usepackage{cleveref}  
\usepackage{amsbsy}
\usepackage{bm}
\usepackage{url}
\usepackage{color}
\usepackage{multirow}
\usepackage{booktabs}
\usepackage{autobreak}
\allowdisplaybreaks 

\newcommand{\inlineTag}{
    \refstepcounter{equation}
    \bgroup\normalfont\normalcolor (\theequation)\egroup}

\crefname{equation}{Eq.}{Eqs.}%
\crefname{figure}{Fig.}{Figs.}%

\usepackage{algorithm}
\usepackage[noend]{algpseudocode}

\algnewcommand{\Break}{\textbf{break}}
\algrenewcommand\algorithmicindent{1em}

\everydisplay{\small} 

\newcommand\Erase{\bgroup\markoverwith{\textcolor{red}{\rule[.5ex]{2pt}{0.4pt}}}\ULon}

\begin{document}

    \title{
        On Neural Network Identification for Low-Speed Ship Maneuvering Model
    }
    
    \author{
        Kouki Wakita$^{1}$ \and
        Atsuo Maki$^{1}$ \and
        Umeda Naoya$^{1}$ \and
        Yoshiki Miyauchi$^{1}$ \and
        Tohga Shimoji$^{1}$ \and
        Dimas M. Rachman$^{1}$ \and
        Youhei Akimoto$^{2,3}$
    }

    \date{%
    \flushleft{\footnotesize
        $^1$Osaka University, 2-1 Yamadaoka, Suita, Osaka, Japan \\%
        $^2$Faculty of Engineering, Information and Systems, University of Tsukuba, 1-1-1 Tennodai, Tsukuba, Ibaraki 305-8573, Japan \\
        $^3$RIKEN Center for Advanced Intelligence Project, 1-4-1 Nihonbashi, Chuo-ku, Tokyo 103-0027, Japan \\[2ex]%
        Email: kouki\_wakita@naoe.eng.osaka-u.ac.jp; maki@naoe.eng.osaka-u.ac.jp \\
    }}


    \maketitle

    \begin{abstract}
        Several studies on ship maneuvering models have been conducted using captive model tests or computational fluid dynamics (CFD) and physical models, such as the maneuvering modeling group (MMG) model. A new system identification method for generating a low-speed maneuvering model using recurrent neural networks (RNNs) and free running model tests is proposed in this study. We especially focus on a low-speed maneuver such as the final phase in berthing to achieve automatic berthing control. Accurate dynamic modeling with minimum modeling error is highly desired to establish a model-based control system. We propose a new loss function that reduces the effect of the noise included in the training data. Besides, we revealed the following facts --- an RNN that ignores the memory before a certain time improved the prediction accuracy compared with the ``standard'' RNN, and the random maneuver test was effective in obtaining an accurate berthing maneuver model. In addition, several low-speed free running model tests were performed for the scale model of the M.V. Esso Osaka. As a result, this paper showed that the proposed method using a neural network model could accurately represent low-speed maneuvering motions.
    \end{abstract}

    \section{Introduction}\label{Sec-1}
        In Japan, a shortage in human resources has recently become a critical issue in coastal shipping due to the aging of ship crews \cite{whitepaper2020}. Therefore, the autonomization in domestic shipping is required. To realize this autonomization, automating berthing control is one of the challenges. So far, various studies have been conducted on the automatic berthing problem \cite{Kose1984,yamato1990,Shouji1992,hasegawa1993,Maki2020,Miyauchi2021}. In addition, we have conducted studies of offline berthing path planning \cite{Maki2020,Miyauchi2021} and online berthing controller \cite{Dimas2020,Dimas2021,Akimoto2021,wakita2021}. Since these control methods are model-based controls, their performances can depend on the modeling error of the given state equations. Although the modeling error should be absorbed by the online feedback control method, the smaller the modeling error, the more realistic the berthing path planning and online control can be achieved.
        
        Recently, some significant studies on developing ship maneuvering models have been conducted \cite{Abkowitz1980,yasukawa2015}. The widely used simulation models are based on the outcome of hydrodynamics, and they can be represented by a set of equations of motion with several coefficients to be determined by captive model tests \cite{yasukawa2015} and computational fluid dynamics (CFD) \cite{SAKAMOTO2019}. Dynamical system models can be categorized into two --- \textit{white-box} models, where the systems are usually understandable and explicitly described as a form of ``standard'' equations, and \textit{black-box} models, where the systems do not necessarily have an explicit form of ``standard'' equations of motion and usually incomprehensible. 
    
        Examples of the white-box maneuvering models are the MMG model \cite{yasukawa2015} mainly developed in Japan and the Abkowitz model \cite{Abkowitz1980}; they are commonly used in practice. The coefficients included in these models can be determined through several captive model tests or CFD. Determining the coefficients by the captive model tests in the tanks use common techniques, such as circular motion test (CMT), planer motion mechanism (PMM) test, and so on. One problem is the difference in the Reynolds number between the scale model and full-scale ship. One of the final goals of modeling is to obtain the maneuvering model in full scale to achieve automatic berthing. Therefore, the difference in Reynolds number can deteriorate modeling accuracy in a full-scale ship. Although it can be possible to conduct full-scale computation in CFD, it requires an extremely high computational cost at present. Therefore, system identification (SI) from the actual ship maneuvers is considered a practical method \cite{SUTULO2014}. So far, several studies on SI for maneuvering models have been conducted \cite{Abkowitz1980,ARAKI2012,Miyauchi2020,MUNOZMANSILLA2009,Nagumo1967,KALLSTROM1981,ASTROM1980,Perera2015}. 
        
        Despite the abovementioned advantage, SI with white-box models has several drawbacks. One of the ways to enhance the performance of at white-box model is to increase the number of the component inside the model. Particularly, for the MMG model \cite{yasukawa2015}, enhancing the modeling based on hydrodynamic insight is essential. In a low-speed berthing maneuver, the flow field around a vessel is apparently complex, so its hydrodynamic modeling is complicated. For instance, in the conventional MMG model, the included parameters, which must implicitly depend on propeller slip ratio or ship speed, e.g., steering resistance reduction factor $t_R$, rudder force increase factor $a_H$, are often considered as constant \cite{yasukawa2015, Yasukawa2021}. Moreover, it is essential to switch a hydrodynamic model depending on the direction of the longitudinal speed and propeller revolution based on tank test results and hydrodynamic insight, which means using numerous conditional branches, i.e., \texttt{if} sentences. Therefore, these efforts seem difficult, and it is time-consuming. To achieve these, the dynamical system model becomes more complex. Thus, black-box models are considered suitable for such cases. 
    
        Many studies have been conducted on SI for ship maneuvers using black-box models, such as support vector machine \cite{XU2020,LUO2014,Luo2013SVM,ZHU2020}, random forest \cite{Mei2019}, and neural networks (NNs) \cite{MOREIRA2003,OSKIN2013,Chiu2004,RAJESH2008,Luo2016NN,Koda2020-1,Koda2020-2}. In particular, an NN can approximate various functions using a set of the appropriate number of parameters, as shown in the universal approximation theorem (UAT) \cite{FUNAHASHI1989183,Cybenko1989,HORNIK1991251}, and the gradient can easily be calculated using the backpropagation (BP) method \cite{rumelhart1985}. Among NNs, a recurrent NN (RNN), which considers memory, seems to be suitable for a maneuvering model prediction problem that is categorized into the partially observable system. Moreover, several studies \cite{MOREIRA2003,OSKIN2013,Chiu2004,Koda2020-1,Koda2020-2}  have shown that the RNN could be a practical method for predicting maneuvering models.
        
        However, in the abovementioned studies, the following issues are still left as an open problem. The first issue is the choice of the network structure. In previous studies, two types of ``RNN'' were used --- the ``standard'' RNN which was used in \cite{MOREIRA2003,Chiu2004,Koda2020-1,Koda2020-2}, and the RNN that ignores the memory before a certain time, which was used \cite{OSKIN2013}. In this study, although it is unclear whether the employed latter NN is an RNN, we treat it as RNN. Both types of RNN are used for the ship maneuvering model. To the best of our knowledge, no study directly compares these two types of RNN.
        
        The second issue is the objective function in the training process. In the continuous control problem of ship maneuver, the right side of the state equation outputs the instantaneous accelerations with respect to each instantaneous state variable, such as velocities and yaw angular velocity. Therefore, the dynamical system modeling employs an NN is expected to output instantaneous accelerations similar to the conventional ship maneuvering mathematical model. In most previous studies that use NN for the prediction of ship maneuvering dynamical system modeling, instantaneous accelerations, which are outputs of the NN, were directly evaluated by loss function in the optimization process.
        This evaluation process is considered appropriate for numerically generated training data. However, training data measured in a real environment contain enormous observation noise and uncertainties in the acceleration component. 
        For instance, in our experimental system, accelerations were obtained from the time-differentiation of positions $X$ and $Y$ measured by the global navigation satellite system (GNSS), and apparently the derivative operations are likely to induce noise. Therefore, it is not easy to predict a ship maneuvering model in the actual uncertain environment using existing methods.
        
        The third issue is the target speed and maneuver range of a ship maneuvering model. In previous studies, NNs were not trained and tested for complex maneuvers, such as the final phase maneuvering motion in berthing control. To train an NN model for low-speed manuevering, both combinations of positive/negative propeller revolution per second (RPS) and ship-positive/negative velocity should be included in the training datasets.
        In almost all SI studies on ship maneuvering models using NNs  \cite{Chiu2004,RAJESH2008,Luo2016NN,MOREIRA2003}, the target is the maneuvering motion with positive speed and positive propeller revolution number, and the maneuver scenarios are mainly ``usual maneuver,'' such as zigzag and turning tests. In \cite{OSKIN2013}, the propeller revolution number was constant.
        However, if the target is a low-speed maneuver, such as the berthing maneuver, 
        the training dataset that only comprises zigzag and turning tests are insufficient, and not only the NN structure but also the acquisition procedure of the training data itself should be analogously reconsidered. This is because to successfully proceed with the training for a low-speed maneuver, it is unavoidable to correct a ``better'' dataset, including positive/negative propeller RPS and forward/backward longitudinal speeds. Therefore, it is desired that  the combination of states and these inputs should be uniformly distributed. 
        
        The major contributions of this article are as follows. (i) Proposing a methodology to predict the dynamical system modeling using an NN that is applicable for a low-speed maneuver, such as berthing maneuver. (ii) Revealing that the RNN that ignores the memory before a certain time improves the prediction accuracy compared with the ``standard'' RNN.
        (iii) Proposeing a new loss function to reduce the effect of noise using numerical simulation. (iv) Revealing the effectiveness of random maneuvering tests to obtain more uniformly distributed combinations of state and input used when a berthing maneuver is predicted using NN.
        
        The rest of this article is organized as follows. Section \ref{Sec-2} describes the problem settings, an NN architecture, the newly introduced loss function, and a white-box model used for validation --- an MMG model. Section \ref{Sec-3} describes the subject ship and training data acquisition in our experimental pond. In Section \ref{Sec-4}, the states and trajectories calculated from the trained black-box model are presented, and the prediction performances of the emplayed NN and MMG model are compared for validation. The dependency of the prediction accuracy on the amount of the data is also discussed. Finally, Section 5 concludes this study.
        
    \section{Mathematical Modeling with NN}\label{Sec-2}
        In this section, we explain the methodology to estimate the maneuvering model using an NN model. 
        
\begin{figure}[htbp]
            \centering
            \includegraphics[width=\columnwidth]{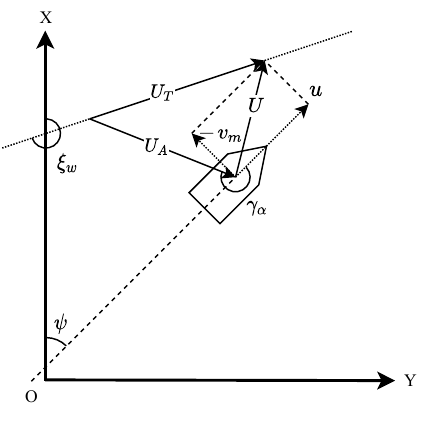}
            \caption{Coordinate Systems.}
            \label{fig:coordinate_systems}
        \end{figure}

        Two coordinate systems are used (\Cref{fig:coordinate_systems}). One is an earth-fixed coordinate system $\mathrm{O}-XY$, and the other is a ship-fixed coordinate system $\mathrm{O}-xy$, with the midship as the origin.
    
        A state equation of maneuvering motion assuming the Markov property is described as follows. The inputs of the state equation are the state variable, control variables, and external disturbance. Concerning the external disturbance, the wind is assumed to be measurable. Other disturbances, such as tide, and vortices may also affect the ship’s maneuverability, but these properties are regarded as unknown. Meanwhile, the outputs of the state equation are instantaneous acceleration components. Summarizing the above, the usual state equation of maneuvering motion, which is the Markov process is as follows:
        \begin{equation}
            \dot{\bm{x}} = \bm{f}\left( \,\bm{x},\, \bm{u},\, \bm{w}\, \right) \enspace ,
            \label{eq:dynamics}
        \end{equation}
        where the overdot $\dot{ }$ denotes the derivative with respect to time $t$. The state vector $\bm{x} \in \mathbb{R}^6$, the control vector $\bm{u} \in \mathbb{R}^2$, and the wind disturbance vector $\bm{w} \in \mathbb{R}^2$ were defined as follows,
        \begin{equation}
            \left\{
            \begin{aligned}
                \bm{x} &\equiv \left(\, X, \enspace u,\enspace Y, \enspace v_m, \enspace \psi, \enspace r \, \right)^{\mathrm{T}}\\
                \bm{u} &\equiv \left(\, n, \enspace \delta \,\right)^{\mathrm{T}}\\
                \bm{w} &\equiv \left(\, U_{A}, \enspace \gamma_{\alpha} \,\right)^{\mathrm{T}} \enspace.
            \end{aligned}
            \right.
            \label{eq:definevariable}
        \end{equation}
        Here, $X\,[\,\mathrm{m}\,]$ and $Y\,[\,\mathrm{m}\,]$ are the ship’s position in the earth-fixed coordinate system. The heading (yaw) $\psi\,[\,\mathrm{rad}\,]$ is the angle between the earth-fixed coordinate system and the ship-fixed coordinate system. $u\,[\,\mathrm{m/s}\,]$ and $v\,[\,\mathrm{m/s}\,]$ are the longitudinal (surge) velocity and the lateral (sway) velocity at the center of gravity, respectively. These velocities are defined in the ship-fixed coordinate system. The yaw angular velocity is denoted by $r\,[\,\mathrm{1/s}\,]$. The relation between $v$ and $v_m\,[\,\mathrm{m/s}\,]$ is defined as follows:
        \begin{equation}
            v_m = v - x_G r \enspace,
        \end{equation}
        where $v_m$ and $x_G$ are the sway velocity at the midship and the distance of the center of gravity from the midship, respectively. The control inputs are the propeller RPS $n\,[\,\mathrm{1/s}\,] \, \rm{or} \, \,[\,\mathrm{rps}\,]$ and the rudder angle $\delta\,[\,\mathrm{rad}\,]$. Meanwhile, the wind disturbance vector comprises of the apparent wind speed $U_{A}\,[\,\mathrm{m/s}\,]$ and the apparent wind direction $\gamma_{\alpha}\,[\,\mathrm{rad}\,]$. These apparent wind properties, i.e., $U_{A}\,[\,\mathrm{m/s}\,]$ and $\gamma_{\alpha}\,[\,\mathrm{rad}\,]$, are calculated from the absolute wind speed $U_\mathrm{T}\,[\,\mathrm{m/s}\,]$, absolute wind direction $\xi_{w}\,[\,\mathrm{rad}\,]$, longitudinal velocity $u$, and the lateral velocity $v_m$ as shown in \cref{fig:coordinate_systems}
        
        Besides, for convenience, the state variables $\bm{x}$ were categorized into two --- position and velocity components --- as follows:
        \begin{equation}
            \left\{
            \begin{aligned}
                \bm{p} &\equiv \left(\, X, \enspace Y, \enspace \psi \, \right)^{\mathrm{T}}\\
                \bm{v} &\equiv \left(\, u, \enspace v_m, \enspace r \, \right)^{\mathrm{T}} \enspace.
            \end{aligned}
            \right.
            \label{eq:statesplit}
        \end{equation}
        Then, the following relationship between $\left(\, X,\, Y,\, \psi \, \right)^{\mathrm{T}}$ and $\left(\,u,\, v_m,\, r \, \right)^{\mathrm{T}}$ should be satisfied:
        \begin{equation}
            \frac{\mathrm{d}}{\mathrm{d}t}
            \left(
                \begin{array}{c}
                    X \\
                    Y \\
                    \psi \\
                \end{array}
            \right)
            = \left(
            \begin{array}{c}
                u \cos{\psi} - v_m \sin{\psi} \\
                u \sin{\psi} + v_m \cos{\psi} \\
                r \\
            \end{array}
            \right) \enspace .
        \end{equation}
        Using the above relationship, it is easy to predict the derivative of the position component $\bm{p}$ in \Cref{eq:dynamics}. Therefore, only the derivative of the velocity component $\bm{v}$, namely acceleration, as shown in the underlined part in the following equation, is the target to be predicted:
        \begin{equation}
            \begin{array}{lll}
            \dot{\bm{x}} = \left(
                    \begin{array}{c}
                        u \cos{\psi} - v_m \sin{\psi} \\
                        \underline{ \dot{u}\left(\, \bm{x},\, \bm{u},\, \bm{w}\, \right) } \\
                        u \sin{\psi} + v_m \cos{\psi} \\
                        \underline{ \dot{v}_{m}\left(\, \bm{x},\, \bm{u},\, \bm{w}\, \right) } \\
                        r \\
                        \underline{ \dot{r}\left(\, \bm{x},\, \bm{u},\, \bm{w}\, \right) }
                    \end{array}
                \right) \enspace.
            \end{array}
            \label{eq:estterm}
        \end{equation}
        
        Up to this point, for simplicity, we have assumed that the state equation has the Markov property. However, the state equation often violates the Markov property. In such a situation, the output of the state equation depends also on the past, i.e., $\dot{\bm{x}}$ at time $\tau$ depends on $\bm{x}_{t \leq \tau}$, $\bm{u}_{t \leq \tau}$, and $\bm{w}_{t \leq \tau}$. Hence, our target is to estimate $\dot{u}$, $\dot{v}_m$, $\dot{u}$ in \Cref{eq:estterm} using the NN, where their inputs are replaced with $\bm{x}_{t \leq \tau}$, $\bm{u}_{t \leq \tau}$, and $\bm{w}_{t \leq \tau}$. The first, third and fifth elements of the right-hand side of \Cref{eq:estterm} remain unchanged and we need not estimate them.
        

        \subsection{NN Model}
            A artificial NN \cite{rumelhart1985,rosenblatt1958} is a computational model that imitates the NN of the human brain. As already mentioned, an NN can represent various functions using the set of appropriate parameters, as shown in the UAT \cite{FUNAHASHI1989183,Cybenko1989,HORNIK1991251}, and the gradient can be easily calculated using the BP method \cite{rumelhart1985}. NNs are widely used in many engineering fields, such as image recognition \cite{Krizhevsky2012}, language processing \cite{Vinyals2015}, and anomaly detection \cite{Raghavendra2018}. Considering these advantages of NNs, we conducted a study on SI for ship maneuver using an NN.
            
            As stated above, in addition to the state variable $\bm{x}$, the control variable $\bm{u}$, and the wind disturbance $\bm{w}$, there are several unknown, uncertain, and unobservable factors such as the vortices around the hull. The vortices induce the memory effect, so the state equation is no more a Markov process. In this sense, the RNN, which considers memory, is appropriate for this partially observable and non-Markov system. From such a perspective, some studies on the prediction of maneuvering models using RNNs have been conducted \cite{MOREIRA2003,OSKIN2013,Chiu2004}, and their effectiveness has been shown. Therefore, we apply the RNN to a low-speed maneuvering prediction problem.
            
            In this study, we assumed that the acceleration vector $\bm{a} \equiv (\dot{u}, \dot{v}_m, \dot{r})$, which is predicted as an output of the NN, is independent of the positions and yaw angle vector $\bm{p}$. Therefore, $\bm{v}$, $\bm{u}$, and $\bm{w}^{\prime}$ are considered the input features. Concerning $\bm{w}^{\prime}$ in the input, directly using the apparent wind direction $\gamma_{\alpha}$ is unnecessarily appropriate for an input feature because of its sudden jump of value at $0$ or $2 \pi$. The apparent wind speed and direction $(U_{A},\gamma_{\alpha})$ are converted into the apparent wind speed vector $\bm{w}^{\prime}$ on the ship-fixed coordinate system as follows:
            \begin{equation}
                \bm{w}^{\prime} \equiv \left(\, U_{A}\cos \gamma_{\alpha}, \enspace U_{A}\sin \gamma_{\alpha} \, \right)^{\mathrm{T}} \enspace.
                \label{eq:windvector}
            \end{equation}
            
            As mentioned above, two types of RNNs have been used in previous studies. One is the ``standard'' RNN \cite{MOREIRA2003,Chiu2004,Koda2020-1,Koda2020-2}, which considers memory by substituting the output of a previous step into the input of the present step; notably, all previous memories are considered. The other one is the RNN \cite{OSKIN2013} in which a limited number of previous memory is explicitly used. In this RNN, the state variables in the recent past are considered, and the states in the distant past are ignored. Generally, states of the recent past strongly affect the present motion. From such a perspective, explicitly considering the recent past states could be beneficial. Therefore, in this study, both models are employed and compared.  
            
            We now explain two types of networks used in this paper. First, the ``standard'' RNN is formulated as follows:
            \begin{equation}
                \left\{\begin{aligned}
                	&\bm{z}_{1}(t_{i}) = \tanh(\bm{W}^{\mathrm{T}}_{x0}\bm{v}(t_{i}) + \bm{W}^{\mathrm{T}}_{u0}\bm{u}(t_{i}) + \\ & \quad \quad \quad \quad \bm{W}^{\mathrm{T}}_{w0}\bm{w}^{\prime}(t_{i}) + \bm{W}^{\mathrm{T}}_{r0}\bm{z}_{1}(t_{i-1}) + \bm{b}_{0}) - \tanh(\bm{b}_{0}) \\
                	&\bm{z}_{2}(t_{i}) = \tanh(\bm{W}^{\mathrm{T}}_{1}\bm{z}_{1}(t_{i})  + \bm{b}_{1}) - \tanh(\bm{b}_{1}) \\
                	&\bm{z}_{3}(t_{i}) = \tanh(\bm{W}^{\mathrm{T}}_{2}\bm{z}_{2}(t_{i})  + \bm{b}_{2}) - \tanh(\bm{b}_{2}) \\
                	&\bm{a}_{NN}(t_{i}) = \bm{W}^{\mathrm{T}}_{3}\bm{z}_{3}(t_{i}) \enspace.
                \end{aligned}\right.
                \label{eq:nnmodel1}
            \end{equation}
            where, $t_{i}$ is the discretized time and is given by $t_{i} = t_{0} + i \Delta t, (i = 0,1,2 \ldots)$, with $t_{0}$: initial time; $\Delta t$: sample period; $\bm{a}_{NN}$ is the output vector predicting $\bm{a}$; $\bm{z}_{1}$,$\bm{z}_{2}$ and $\bm{z}_{3}$ are the vectors of latent variables and $\bm{z}_1(t_0) = \bm{0}$. $\bm{W}_{x0} \in \mathbb{R}^{3 \times 200}$, $\bm{W}_{u0} \in \mathbb{R}^{2 \times 200}$, $\bm{W}_{w0} \in \mathbb{R}^{2 \times 200}$, $\bm{W}_{r0} \in \mathbb{R}^{200 \times 200}$, $\bm{W}_{1} \in \mathbb{R}^{200 \times 200}$, $\bm{W}_{2} \in \mathbb{R}^{200 \times 200}$, $\bm{W}_{3} \in \mathbb{R}^{200 \times 200}$, $\bm{b}_{0} \in \mathbb{R}^{200}$, $\bm{b}_{1} \in \mathbb{R}^{200}$, and $\bm{b}_{2} \in \mathbb{R}^{200}$ are the parameters included in the model; these parameters are represented collectively as $\bm{\theta}$. Formally, the input-output relation at time $t_i$ is expressed as follows:
            \begin{equation}
                \bm{a}_{NN}(t_i) = \bm{f}_{1NN}( \bm{v}(t_i), \bm{u}(t_i), \bm{w}'(t_i), \bm{z}_1(t_{i-1}); \bm{\theta} ) \enspace,
            \end{equation}
            where $\bm{z}_1(t_{i-1})$ plays the role of the memory representing the past, i.e., $\bm{v}(t_{j})$, $\bm{u}(t_{j})$ and $\bm{w}'(t_{j})$ for $j < i$.
            
            The second terms of first, second, and third expressions in \Cref{eq:nnmodel1}, $\tanh(\bm{b}_{k})$ for $k=0,1,2$, are subtracted to impose zero output on the NN model at the origin of input features. This type of NN model was introduced in \cite{Nakanishi1997}. The constraints enable us to consider that the accelerations must be zero in principle without the existence of external forces and vessel speeds.
            
            Second, the NN that explicitly considers the recent past memories of states variables can be expressed as follows:
            \begin{equation}
                \left\{\begin{aligned}
                	&\bm{z}_{1}(t_{i-m+1}) = \tanh(\bm{W}^{\mathrm{T}}_{x0}\bm{v}(t_{i-m+1}) + \bm{W}^{\mathrm{T}}_{u0}\bm{u}(t_{i-m+1}) + \\ & \quad \quad \quad \quad \bm{W}^{\mathrm{T}}_{w0}\bm{w}^{\prime}(t_{i-m+1}) + \bm{b}_{0}) - \tanh(\bm{b}_{0}) \\
                	&\bm{z}_{1}(t_{i-m+2}) = \tanh(\bm{W}^{\mathrm{T}}_{x0}\bm{v}(t_{i-m+2}) + \bm{W}^{\mathrm{T}}_{u0}\bm{u}(t_{i-m+2}) + \\ &  \quad \quad \bm{W}^{\mathrm{T}}_{w0}\bm{w}^{\prime}(t_{i-m+2}) + \bm{W}^{\mathrm{T}}_{r0}\bm{z}_{1}(t_{i-m+1}) + \bm{b}_{0}) - \tanh(\bm{b}_{0}) \\
                	& \quad \quad  \quad \quad \vdots \\
                	&\bm{z}_{1}(t_{i}) = \tanh(\bm{W}^{\mathrm{T}}_{x0}\bm{v}(t_{i}) + \bm{W}^{\mathrm{T}}_{u0}\bm{u}(t_{i}) + \\ & \quad \quad \quad \quad \bm{W}^{\mathrm{T}}_{w0}\bm{w}^{\prime}(t_{i}) + \bm{W}^{\mathrm{T}}_{r0}\bm{z}_{1}(t_{i-1}) + \bm{b}_{0}) - \tanh(\bm{b}_{0}) \\
                	&\bm{z}_{2}(t_{i}) = \tanh(\bm{W}^{\mathrm{T}}_{1}\bm{z}_{1}(t_{i})  + \bm{b}_{1}) - \tanh(\bm{b}_{1}) \\
                	&\bm{z}_{3}(t_{i}) = \tanh(\bm{W}^{\mathrm{T}}_{2}\bm{z}_{2}(t_{i})  + \bm{b}_{2}) - \tanh(\bm{b}_{2}) \\
                	&\bm{a}_{NN}(t_{i}) = \bm{W}^{\mathrm{T}}_{3}\bm{z}_{3}(t_{i}) \enspace.
                \end{aligned}\right.
                \label{eq:nnmodel2}
            \end{equation}
            The parameters of \Cref{eq:nnmodel2} are equivalent to those of \Cref{eq:nnmodel1}. The difference between \Cref{eq:nnmodel1} and (\ref{eq:nnmodel2}) is that \Cref{eq:nnmodel2} ignores the memory before $t_{i-m+1}$ by explicitly substituting the input features of $t_{i-j}$ for $j=0,1,\ldots,m-1$. \Cref{eq:nnmodel1} can be understood as the limit as $m \rightarrow \infty$ in \Cref{eq:nnmodel2}.
            
            Formally, the input-output relation at time $t_i$ is expressed as follows:
            \begin{equation}
                \begin{aligned}
                	\bm{a}_{NN}(t_{i}) = \bm{f}_{2NN}( 
                	&\bm{v}(t_{i}), \bm{v}(t_{i-1}), \cdots ,\bm{v}(t_{i-m+1}),\\ 
                	&\bm{u}(t_{i}), \bm{u}(t_{i-1}), \cdots, \bm{u}(t_{i-m+1}),\\
                	&\bm{w}^{\prime}(t_{i}), \bm{w}^{\prime}(t_{i-1}), \cdots, \bm{w}^{\prime}(t_{i-m+1}); \bm{\theta}) \enspace,
                \end{aligned}
            \end{equation}
            
            The prediction accuracies of the two models will be compared in Section \ref{Sec-4}.
            
            \begin{table*}[htbp]
                \scriptsize
                \centering
                \caption{
                    Training process. Here, acceleration components are directly evaluated. In this table, both formulations for \cref{eq:nnmodel1} or \cref{eq:nnmodel2} are listed. 
                }
                \begin{tabular}{c||l|l} \hline
                    & The case of \cref{eq:nnmodel1} & The case of \cref{eq:nnmodel2} \\ \hline \hline
                    Required Dataset & \multicolumn{2}{c}{
                        $\left\{ \left\{\hat{\bm{x}}(t_{n,i}), \hat{\bm{u}}\left(t_{n,i}\right), \hat{\bm{w}}^{\prime}\left(t_{n,i}\right), \hat{\bm{a}}\left(t_{n,i}\right)\right\}_{i=0, \ldots, N_{T}-1} \right\}_{n = 1 \ldots, N} \quad \inlineTag\label{eq:dataset1}$
                    } \\ 
                    Initialization & $\bm{z}_{1}(t_{0}) = \bm{0} \quad \inlineTag\label{eq:initacc1}$ & $\bm{a}_{NN}(t_{n,j}) = \hat{\bm{a}}_{NN}(t_{n,j})  \quad \text{for} \quad j = 0, \ldots, m-1 \quad \inlineTag\label{eq:initacc2}$ \\ 
                    Iteration Process & $
                        \begin{aligned}
                        	&\bm{a}_{NN}(t_{n,i}) = \\
                        	& \quad \bm{f}_{1NN}( \hat{\bm{v}}(t_{n,i}), \hat{\bm{u}}(t_{n,i}), \hat{\bm{w}}(t_{n,i}), \bm{z}_1(t_{n,i-1}); \bm{\theta} )
                        \end{aligned} \quad \inlineTag\label{eq:predacc1}$ & $\begin{aligned}
                    	\bm{a}_{NN}(t_{n,i}) = \bm{f}_{2NN}(
                    	&\hat{\bm{v}}(t_{n,i}), \hat{\bm{v}}(t_{n,i-1}), \cdots ,\hat{\bm{v}}(t_{n,i-m+1}),\\ 
                    	&\hat{\bm{u}}(t_{n,i}), \hat{\bm{u}}(t_{n,i-1}), \cdots, \hat{\bm{u}}(t_{n,i-m+1}),\\
                    	&\hat{\bm{w}}^{\prime}(t_{n,i}), \hat{\bm{w}}^{\prime}(t_{n,i-1}), \cdots, \hat{\bm{w}}^{\prime}(t_{n,i-m+1}); \bm{\theta})
                    \end{aligned} \quad \inlineTag\label{eq:predacc2}$ \\ 
                    Loss Function & \multicolumn{2}{c}{
                        $\mathcal{L}_{acc}(\bm{\theta})=\frac{1}{N}\frac{1}{N_{T}} \sum_{n=1}^{N}\sum_{i=0}^{N_{T}-1}\sum_{j=1}^{3}\left|\frac{a_{NN,j}(t_{n,i}) - \hat{a}_{j}(t_{n,i})}{\sigma_{\hat{a}_{j}}}\right|^{2} \quad \inlineTag\label{eq:accloss}$
                    } \\
                    \hline
                \end{tabular}
                \label{tab:trainingflowacc}
            \end{table*}
        
            \begin{table*}[htbp]
                \scriptsize
                \centering
                \caption{
                    Training process.  Here, only state variables are evaluated, and acceleration components are not directly evaluated. In this table, both formulations for \cref{eq:nnmodel1} or \cref{eq:nnmodel2} are listed.  
                }
                \begin{tabular}{c||l|l} \hline
                    & The case of \cref{eq:nnmodel1} & The case of \cref{eq:nnmodel2} \\ \hline \hline
                    Required Dataset & \multicolumn{2}{c}{
                        $\left\{ \left\{\hat{\bm{x}}(t_{n,i}), \hat{\bm{u}}\left(t_{n,i}\right), \hat{\bm{w}}^{\prime}\left(t_{n,i}\right)\right\}_{i=0, \ldots, N_{T}-1} \right\}_{n = 1 \ldots, N} \quad \inlineTag\label{eq:dataset2}$
                    } \\
                    Initialization & $\left\{\begin{aligned}
                            \bm{p}(t_{n,0}) &= \hat{\bm{p}}(t_{n,0}) \\
                            \bm{v}(t_{n,0}) &= \hat{\bm{v}}(t_{n,0})
                        \end{aligned}\right. \quad \inlineTag\label{eq:initstate1}$ & $\left\{\begin{aligned}
                            \bm{p}(t_{n,i}) &= \hat{\bm{p}}(t_{n,i}) \\
                            \bm{v}(t_{n,i}) &= \hat{\bm{v}}(t_{n,i})
                        \end{aligned}\right. \quad \text{for} \quad i = 0, \ldots, m-1 \quad \inlineTag\label{eq:initstate2}$ \\ 
                    \begin{tabular}{c} Iteration Process \\ (Position) \end{tabular} & \multicolumn{2}{c}{$\left\{\begin{aligned}
                    &\dot{\bm{p}}(t_{n,i}) =  \left(
                            \begin{array}{c}
                                u(t_{n,i}) \cos{\psi(t_{n,i})} - v_m(t_{n,i}) \sin{\psi(t_{n,i})} \\
                                u(t_{n,i}) \sin{\psi(t_{n,i})} + v_m(t_{n,i}) \cos{\psi(t_{n,i})} \\
                                r(t_{n,i}) \\
                            \end{array}
                            \right)  \\
                            &\bm{p}(t_{n,i+1}) = \bm{p}(t_{n,i}) +  \Delta t \dot{\bm{p}}(t_{n,i}) 
                        \end{aligned}\right. \quad \inlineTag\label{eq:simpos}$} \\
                    \begin{tabular}{c} Iteration Process \\ (Velocity) \end{tabular} & $\left\{\begin{aligned}
                            & \bm{a}_{NN}(t_{n,i}) = \\
                            & \quad \bm{f}_{1NN}(\bm{v}(t_{n,i}), \hat{\bm{u}}(t_{n,i}), \hat{\bm{w}}^{\prime}(t_{n,i}), \bm{z}_1(t_{n,i-1}); \bm{\theta} ) \\
                            & \bm{v}(t_{n,i+1}) = \bm{v}(t_{n,i}) + \Delta t \bm{a}_{NN}(t_{n,i})
                        \end{aligned}\right. \quad \inlineTag\label{eq:simvel1}$ & $\left\{\begin{aligned}
                            &\begin{aligned}
                            	&\bm{a}_{NN}(t_{n,i}) = \\
                            	&\quad \begin{aligned}
                                	\bm{f}_{2NN}(&\bm{v}(t_{n,i}), \bm{v}(t_{n,i-1}), \cdots ,\bm{v}(t_{n,i-m+1}),\\ 
                                    &\hat{\bm{u}}(t_{n,i}), \hat{\bm{u}}(t_{n,i-1}), \cdots, \hat{\bm{u}}(t_{n,i-m+1}),\\
                                	&\hat{\bm{w}}^{\prime}(t_{n,i}), \hat{\bm{w}}^{\prime}(t_{n,i-1}), \cdots, \hat{\bm{w}}^{\prime}(t_{n,i-m+1}); \bm{\theta}) \\
                            	\end{aligned}
                            \end{aligned} \\
                            & \bm{v}(t_{n,i+1}) = \bm{v}(t_{n,i}) + \Delta t \bm{a}_{NN}(t_{n,i})
                        \end{aligned}\right. \quad \inlineTag\label{eq:simvel2}$ \\ 
                    Loss Function & \multicolumn{2}{c}{
                        $\mathcal{L}_{state}(\bm{\theta}) = \frac{1}{N}\frac{1}{N_{T}} \sum_{n=1}^{N} \sum_{i=0}^{N_{T}-1} \sum_{j=1}^{6} \left|\frac{x_{j}(t_{n,i}) - \hat{x}_{j}(t_{n,i})}{\sigma_{\hat{x}_{j}}}\right|^{2} \quad \inlineTag\label{eq:stateloss}$
                    } \\
                    \hline
                \end{tabular}
                \label{tab:trainingflowstate}
            \end{table*}

        \subsection{Optimization Methods}
            In previous studies on maneuvering models using NN \cite{MOREIRA2003,Chiu2004,RAJESH2008}, the maneuvering models were predicted by optimizing a loss function in which the instantaneous accelerations were directly evaluated [\Cref{eq:accloss}]. In those existing researches, training data were obtained from numerically simulation with the artificial noises. Therefore, direct evaluation of accelerations was not a fatal issue.
            
            However, the measurement data taken in the actual environment include unignorable noise and uncertainties. In particular, acceleration terms are likely to contain noise, as will be discussed in Section \ref{Sec-3}. In such a case, directly using the measured accelerations in the loss function deteriorates the optimization performance of the NN parameters. 
            
            To handle such a difficulty, in this study, we newly proposed a new loss function to reduce the measurement noise. Thus, as an alternative to directly use acceleration, the loss function evaluates the positions and velocities from numerical simulation. We now explain two loss functions: (I) loss function evaluating acceleration and (I\hspace{-.1em}I) the proposed loss function evaluating state variables, i.e., positions and velocities.
            
            First, we explain (I). The training process is summarized in \Cref{tab:trainingflowacc}. In this method, the trajectory dataset of state variables $\hat{\bm{x}}(t_{n,i})$, control inputs $\hat{\bm{x}}(t_{n,i})$, wind disturbances $\hat{\bm{w}}^{\prime}(t_{n,i})$, and accelerations $\hat{\bm{a}}(t_{n,i})$ are required.
            Subsequently, hat $\, \hat{ } \,$ denotes the training data. $N_{T}$ is the number of time steps of the trajectories, $N$ is the number of these trajectories, and $t_{n,i}$ represents the discretized time of the $i$-th step of the $n$-th trajectory. 
            Then, the acceleration is predicted by the NN model using Eqs.(\ref{eq:initacc1}) and (\ref{eq:predacc1}), or Eqs. (\ref{eq:initacc2}) and (\ref{eq:predacc2}). Thus, the loss function is calculated from the predicted acceleration using \cref{eq:accloss}.
            Here, $a_{NN,j}$ and $\hat{a}_{j}$ represented the $j$-th components of $\bm{a}_{NN}$ and $\hat{\bm{a}}$, respectively. $\sigma_{\hat{a}_{j}}$ is the standard deviation of $\hat{a}_{j}$ in the training dataset. Then, by optimizing this loss function using the gradient descent method, every parameter inside the NN can be determined. In this study, Adam \cite{kingma2014} was used as the optimization method. However, as mentioned above, the measured accelerations $\hat{\bm{a}}$ taken in the actual environment contained an unignorable noise, so applying \cref{eq:accloss} to such noisy data could deteriorate the prediction accuracy.

            Second, we explain (I\hspace{-.1em}I).
            This loss function evaluates the positions and velocities alternative instead of directly evaluating accelerations.
            The training process is shown in \Cref{tab:trainingflowstate}. In this method, the trajectory dataset of state variables $\hat{\bm{x}}(t_{n,i})$, control inputs $\hat{\bm{x}}(t_{n,i})$, and wind disturbance $\hat{\bm{w}}^{\prime}(t_{n,i})$ are required, whereas that of acceleration $\hat{\bm{a}}(t_{n,i})$ is not. To predict the state variables $\bm{x}(t_{n,i})$, numerical simulation is preformed with the measured control inputs $\hat{\bm{u}}(t_{n,i})$, wind disturbances $\hat{\bm{w}}^{\prime}(t_{n,i})$, and initial conditions using \cref{eq:initstate1} or \cref{eq:initstate2}.
            In this study, the Euler method is used to solve ordinary differential equations solver. The prediction of positions and yaw angle $\bm{p}(t_{n,i+1})$ are simulated using \cref{eq:simpos}. Notably, $u(t_{n,i})$, $v_m(t_{n,i})$, and $r(t_{n,i})$ are elements of $\bm{v}(t_{n,i})$ calculated using the Euler method, and $\psi(t_{n,i})$ was an element of $\bm{p}(t_{n,i})$ calculated also using the Euler method. Besides, at the same time, the velocity and angular velocity $\bm{v}(t_{n,i+1})$ can be calculated as \cref{eq:simvel1} or \cref{eq:simvel2}. At this time, the acceleration $\bm{a}_{NN}(t_{n,i})$ is calculated by \cref{eq:nnmodel1} or \cref{eq:nnmodel2}. Notably, the state variables $\bm{v}(t_{n,i})$ are obtained from the simulation, whereas the control input $\hat{\bm{u}}(t_{n,i})$ and wind disturbance $\hat{\bm{w}}^{\prime}(t_{n,i})$ are from the given training data set. Therefore, the position and velocity components can be calculated from the outputs of NN. Finally, the loss function is defined as \cref{eq:stateloss}, where, $x_{j}$ and $\hat{x}_{j}$ represent $j$-th components of $\bm{x}$ and $\hat{\bm{x}}$, respectively. $\sigma_{\hat{x}_{j}}$ is the standard deviation of $\hat{x}_{j}$ in the training data set. Notably that $\bm{x}(t_{n,i})$ comprises $\bm{p}(t_{n,i})$ and $\bm{v}(t_{n,i})$. Then, by optimizing this loss function using the gradient descent method, every parameter inside the NN can be determined.
            
            By introducing the numerical simulation, the direct use acceleration was avoided in the loss function, thus a higher optimization performance can be expected. Analogously, the accumulation error in the simulation can be considered in the loss function. Therefore, using this loss function, it is also expected that the noise effect is relatively reduced. Moreover, as easily imagined, the computational cost due to numerical simulation becomes higher. However, since this calculation is usually performed not online but offline, the cost does not seem to be a critical issue. Further to increase the computational speed, a graphics processing unit (GPU) is used.
            
            The comparison of the two loss functions will be shown in the next Section \ref{Sec-4}. 
        
        \subsection{MMG Model}\label{sec:mmg}
        
            In this article, the MMG model \cite{yasukawa2015} is used for comparison. A brief description of the MMG model used in study is given here. 
    
            First, the equation of maneuvering motion was defined as \Cref{eq:MMGdynamics} based on the MMG model concept:
            \begin{equation}
                \left\{ \begin{array}{ll}
                (m + m_x) \dot{u} - (m + m_y) v_m r -x_{G}mr^{2}= X_H + X_P + X_R \\
                (m + m_x) \dot{v}_m + (m + m_y) u r +x_{G}m\dot{r}= Y_H + Y_P + Y_R \\
                (I_{zz} + J_{zz}+x_{G}^{2}m) \dot{r} + x_G m (\dot{v}_m + ur) = N_H + N_P + N_R \enspace.
                \end{array} \right.
                \label{eq:MMGdynamics}
            \end{equation}
            where, $m$, $m_x$, and $m_y$ represent the ship mass, longitudinal added mass and lateral added mass, respectively; $I_{zz}$ and $J_{zz}$ represent the moment of inertia and added moment of inertia, respectively; and $x_G$ represents the longitudinal position of the center of gravity. On the right-hand side of the equations, $X$, $Y$, and $N$ represent the longitudinal force, lateral force, and yawing moment, respectively. The subscripts H, P, and R represent the hull, propeller, and rudder components, respectively.
            
            The hydrodynamic forces acting on the hull are predicted using a unified model for low-speed maneuvering and open sea navigation \cite{Yoshimura2009}, which is widely used in cases with large oblique angles. For the positive longitudinal speed and positive propeller RPS, the mathematical modeling of the propeller and rudder were based on the conventional MMG model \cite{yasukawa2015}. Meanwhile, for the negative propeller RPS, the models of \cite{Yasukawa2008} for thrust,  \cite{Hachii2004} lateral force, moment by a propeller, and rudder force \cite{Kitagawa2015} are used. For further details of applied models, refer to \cite{Miyauchi20201SI}.

    \section{Subject Ship and Experiment}\label{Sec-3}
\begin{figure}[t]
            \centering
            \includegraphics[width=0.7\columnwidth]{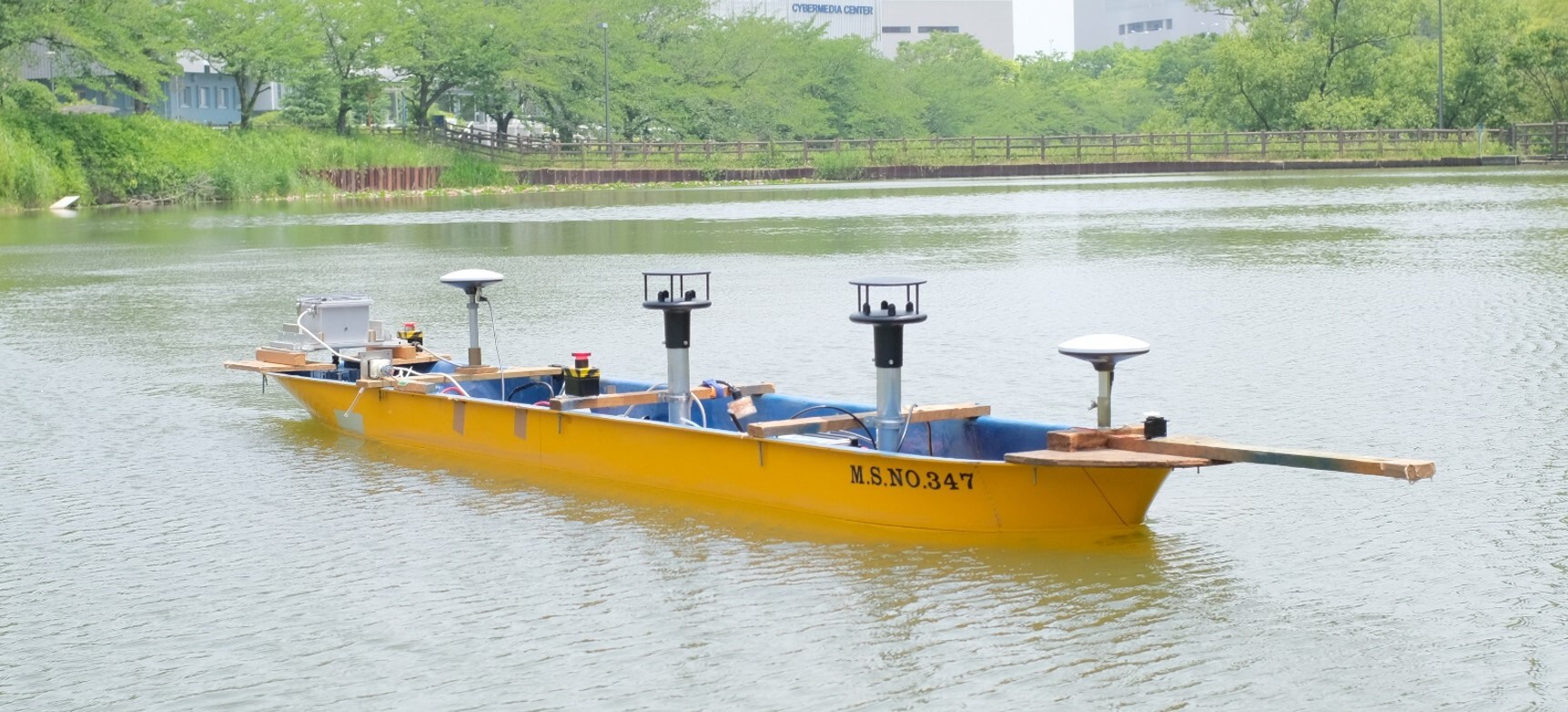}
            \caption{Photo of the subject model ship: M.V. ESSO OSAKA.}
            \label{fig:subjectship}
        \end{figure}

        In this section, the details of free running model tests are explained. The subject model ship was the M.V. ESSO OSAKA (\cref{fig:subjectship}), and the model scale was 1/108.33. The principal particulars of this model are listed in \Cref{tab:principalparticulars}.
        \begin{table}[htbp]
            \centering
            \caption{Principal particulars.}
            \begin{tabular}{ll}
                \hline Item & Value \\
                \hline Length: $L_{p p}$           & $3.0 \mathrm{m}$ \\
                       Breadth: $B$                & $0.49 \mathrm{m}$ \\
                       Draft: $d$                  & $0.20 \mathrm{m}$ \\
                       Block coefficients: $C_{b}$ & $0.83$ \\
                \hline
            \end{tabular}
            \label{tab:principalparticulars}
        \end{table}
        
        The onboard control devices and systems of this model were developed by Wada et al. \cite{Wada2019}. It uses robot operating system (ROS) as the middleware. The upper and lower limits of control variables in the model tests are shown in Table \ref{tab:rangesofcontrolinputs}.
        \begin{table}[htbp]
            \centering
            \caption{Ranges of control inputs}
            \begin{tabular}{ll}
                \hline Item & Range \\
                \hline $n \mathrm{[rps]}$       & $[-20,20]$ \\
                       $\delta \mathrm{[deg.]}$ & $[-35,35]$ \\
                \hline
            \end{tabular}
            \label{tab:rangesofcontrolinputs}
        \end{table}
        
        One optical fiber gyro, three GNSS receivers, and two ultrasonic anemometers were installed in the system. The state variable $\bm{x}$ and disturbance $\bm{w}$ were measured from these measurement devices. The optical fiber gyro directly measured the yaw angular velocity $r$. Although the optical fiber gyro also output $\psi$ as the integration results of $r$, the accumulation of error generated drift of $\psi$. To avoid this drift error, the heading angle $\psi$ was calculated from two or three GNSS data in this experiment. The time series of the control variable $\bm{u}$ were simultaneously recorded. For details of the measurements and analysis methods, see \cite{Miyauchi20201SI}. 
        
        In almost all SI studies on the ship maneuvering models using NN, the target is basically to maneuver with positive longitudinal speed, and maneuver modes are zigzag and turning tests \cite{Chiu2004,RAJESH2008,Luo2016NN,MOREIRA2003}. In \cite{OSKIN2013}, the propeller RPS was constant. However, this study proposed a low-speed berthing maneuvering model that has the form of an NN. In the berthing maneuver, there was a state combination of positive/negative propeller RPSs and the positive/negative longitudinal speeds, hence the model significantly differs from the usual maneuver. Therefore, we handled the training data carefully. As a countermeasures, we conducted random maneuver test to acquire the uniformly distributed combination of states and input.
        
        The random maneuver test is a free running test in which the control input was randomly generated regardless of the state. So, uniformly distributed state and control combinations were expected. In this experiment, to establish the uniformly distributed input and state, manual controls by the gaming controller were conducted. In reality, owing to the limitation of the geometrical size of the pond, the combination did not completely become uniform distribution, but distributed results were obtained \cite{Miyauchi20201SI}. 
        
        As stated above, the turning, zig-zag, random maneuvering, and berthing tests were conducted (\Cref{tab:modeltestlabel}).
        
        \begin{table}[htbp]
            \centering
            \caption{Model tests and its labels}
            \begin{tabular}{llc}
                \hline Label    & Type of Model Test \\ \hline
                \hline T & Turning \\
                       Z & Zigzag \\
                       B & Berthing\,-\,STBD.\,\&\,PORT. \\
                       R & Random \\
                \hline
            \end{tabular}
            \label{tab:modeltestlabel}
        \end{table}
        
        In this study, we prepared one test dataset and three training data sets (\Cref{tab:inputdata}).
        The sampling frequency in each test was $10 \, [ \, \mathrm{Hz} \, ]$. Each test has a label as listed in \cref{tab:modeltestlabel}, and hereafter, for instance, ``TZRB'' in Train-TZRB means that Train-TZRB has turning tests, zig-zag tests, random maneuvering tests, and berthing tests.
        
        \begin{table}[htbp]
            \scriptsize
            \centering
            \caption{List of Training Data Sets. Each sampling frequency was $10 \, [ \, \mathrm{Hz} \, ]$}
            \begin{tabular}{l||lllll}
                \hline Data Set    & T  $\, [ \, \mathrm{s} \, ]$ & Z  $\, [ \, \mathrm{s} \, ]$ & R  $\, [ \, \mathrm{s} \, ]$ & B  $\, [ \, \mathrm{s} \, ]$ & Total $\, [ \, \mathrm{s} \, ]$ \\ \hline
                \hline Train-TZB   & 1490 & 737.1 & 0.0 & 335.8 & 2562.9 \\
                       Train-TZRB  & 556.4 & 342.9 & 1301.2 & 335.8 & 2536.3 \\
                       Train-TZRB+ & 5674.7 & 1151 & 5861.2 & 788.9 & 13475.8 \\
                       Test (TZRB) & 424.6 & 193.8 & 717.9 & 380.6 & 1716.9 \\
                \hline
            \end{tabular}
            \label{tab:inputdata}
        \end{table}
        
        In Train-TZR and Train-TZRB, the amount of training data was intentionally limited to approximately 7 hours in the actual ship scale. Meanwhile, the Train-TZRB+ had almost five times that amount of data. The purpose of using Train-TZRB+ was to confirm the dependency of data quantity.

    \section{Results}\label{Sec-4}
        In this section, the final results and discussion are shown. To confirm the dependency of the prediction accuracy on the NN model, loss function, and data set, the prediction for the maneuvering model was conducted using five methods (\Cref{tab:setting}).
        As will be detailed later, Type-1 has the best performance, so we focus on the explanation of the reason. In Subsection \ref{sec:MDL}, Type-1 and Type-2 are compared. The dependency of prediction accuracy on the NN model is discussed. In Subsection \ref{sec:LM}, Type-1 and Type-3 are compared. We discuss the dependency of prediction accuracy on the loss function. In Subsection \ref{sec:DS}, Type-1, Type-4, and Type-5 are compared. We discuss the dependency of prediction accuracy on the data set. In particular, comparing Type-4 and Type-5, the effectiveness of the random maneuvering test is confirmed. In Subsection \ref{sec:M}, we compare the NN models, particularly Type-1 and Type-5, and the MMG model, whose coefficients are determined by the captive model tests.
        
        To prevent the problem of overfitting, the dataset was divided into two parts --- the training and validation datasets. The training data were used for optimization, whereas the validation data were used to detect the start of overfitting. The parameters inside the NN that minimized the loss function in the validation data were considered the final solution. Python was used as the programming language, and PyTorch was used as the deep learning library. The hyperparameters used in the training are listed in \Cref{tab:hyperparameter}.
        
        \begin{table}[t]
            \scriptsize
            \centering
            \caption{The computational experiments conducted in this study.}
            \begin{tabular}{l||lll} \hline
                & NN Model & Loss Function & Data Set \\ \hline \hline
                Type-1 & \cref{eq:nnmodel2} & \cref{eq:stateloss} & Train-TZRB+ \\
                Type-2 & \cref{eq:nnmodel1} & \cref{eq:stateloss} & Train-TZRB+ \\
                Type-3 & \cref{eq:nnmodel2} & \cref{eq:accloss} & Train-TZRB+ \\
                Type-4 & \cref{eq:nnmodel2} & \cref{eq:stateloss} & Train-TZB \\
                Type-5 & \cref{eq:nnmodel2} & \cref{eq:stateloss} & Train-TZRB \\
                \hline
            \end{tabular}
            \label{tab:setting}
        \end{table}
        
        \begin{table}[t]
            \scriptsize
            \centering
            \caption{Hyperparameter}
            \begin{tabular}{ll}
                \hline Batch size    & $512$ \\
                       Learning rate : $\eta$ & $2.0\times10^{-5}$ or \\
                        &$1.0\times10^{-4}$ ( if loss function is \cref{eq:accloss} ) \\
                       Predicted Steps : $N_{T}$ & $60 \quad ( \, 6.0  \, [ \, \mathrm{s} \, ] \,)$ \\
                       Memory Steps : $m$ & $10 \quad ( \, 1.0  \, [ \, \mathrm{s} \, ] \,)$ \\
                \hline
            \end{tabular}
            \label{tab:hyperparameter}
        \end{table}
        
        \begin{figure}[t]
            \centering
            \includegraphics[width=\columnwidth]{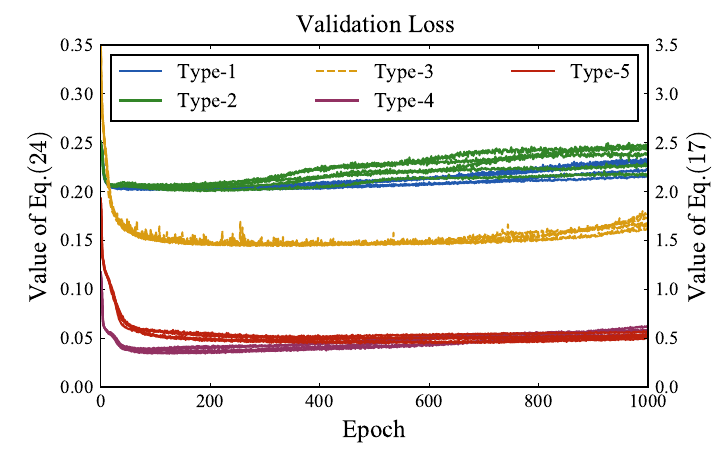}
            \caption{Validation loss value during training.}
            \label{fig:losstransition}
        \end{figure}
        
        \begin{table*}[htbp]
            \scriptsize
            \centering
            \caption{Mean squared error $\mathcal{L}_{MSE}$ of the prediction in Test data set}
            \begin{tabular}{l||l|ll|ll|ll|ll|ll} \hline
                &MMG & Type-1 & & Type-2 & & Type-3 & & Type-4 & & Type-5 & \\ 
                &EDF Coeff.& mean & std & mean & std & mean & std & mean & std & mean & std \\ \hline \hline
                Random    & 0.272 & 0.236 & 0.003 & 0.241 & 0.005 & 4834  & 4041  & 0.488 & 0.022 & 0.262 & 0.007 \\
                Turning   & 0.858 & 0.713 & 0.011 & 0.728 & 0.017 & 0.864 & 0.100 & 0.701 & 0.019 & 0.785 & 0.024 \\
                Zigzag    & 1.015 & 0.086 & 0.063 & 0.124 & 0.141 & 761.1 & 1519  & 0.122 & 0.054 & 0.856 & 0.288 \\
                Berthing1 & 0.036 & 0.052 & 0.024 & 0.058 & 0.031 & 634.8 & 1264  & 0.144 & 0.053 & 0.055 & 0.030 \\
                Berthing2 & 0.189 & 0.069 & 0.014 & 0.140 & 0.101 & 91.76 & 124.6 & 0.246 & 0.107 & 0.182 & 0.047 \\
                Berthing3 & 0.100 & 0.138 & 0.023 & 0.167 & 0.053 & 1.163 & 0.617 & 0.302 & 0.062 & 0.294 & 0.043 \\ \hline
            \end{tabular}
            \label{tab:mse}
        \end{table*}
        
        The training process and prediction results were shown before the comparison of each method. The descent process of the loss functions in the validation dataset during training is shown in \Cref{fig:losstransition}. Generally, optimization of the NN parameters is likely to depend on the initial distribution of parameters, so the training of the NN was performed five times with different random seeds. \Cref{fig:losstransition} shows that the results did not strongly depend on the random seeds in this case. Notably, since the loss functions or the quantity of the datasets were different for each method, the direct comparison of loss value did not have meaning. 
        
        We explain the procedure to evaluate the estimation accuracy of the trained model.
        To quantitatively evaluate the prediction accuracy, the mean squared error (MSE) of the standardized state of trajectories between the prediction results of the NN $\bm{x}$ and experimental results $\hat{\bm{x}}$ was employed. This index was defined as follows:
        \begin{equation}
            \mathcal{L}_{MSE} = \frac{1}{N_{T}} \sum_{i=0}^{N_{T}} \sum_{j=1}^{6} \left|\frac{x_{j}(t_{i}) - \hat{x}_{j}(t_{i})}{\sigma_{\hat{x}_{j}}}\right|^{2}
            \label{eq:mse}
        \end{equation}
        Here, $\bm{\sigma}_{\hat{\bm{x}}}$ is the standard deviation of $\hat{\bm{x}}$ in the test data set.
        
        The comparative results of \cref{eq:mse} between the NN and MMG models are summarized in \Cref{tab:mse}. The results of the NN model were shown as the mean values and standard deviations since the calculations were conducted for five different random seeds.
        The random maneuvering and turning tests had long measurement records. Therefore, to avoid the accumulation of error, the time domain simulation was restarted at every 100 seconds. By avoiding the accumulation of error, we highly expected the optimization performance to improve. At every restart, the initial condition was reset to the measured experimental value.
        Besides, since the setting of initial values in Type-2 was different from others, the way initial values were assigned was changed. While $i=0, 1, \ldots, m-1$, after the simulation was calculated using Eqs. (\ref{eq:simpos}) and (\ref{eq:simvel1}), the position $\bm{p}(t_{n,i})$ and velocity $\bm{v}(t_{n,i})$ were overwritten by \cref{eq:initstate2}. At this time, the latent variable $\bm{z}_{1}$ was not overwritten, but was resubstituted as an input feature in the next time step.
        
        \subsection{Comparison of NN Models}\label{sec:MDL}
            In this section, by comparing Type-1 and Type-2, the dependency of the prediction accuracy on the model was discussed. In Type-1, the RNN that ignores the memory before a certain time expressed in \cref{eq:nnmodel2} was used. In Type-2, the ``standard'' RNN model expressed in \cref{eq:nnmodel1} was used. \cref{eq:stateloss} was used as the loss function, and Train-TZRB+ was used as the training data set.
            
            \Cref{tab:mse} shows that the value $\mathcal{L}_{MSE}$ of Type-1 was always lower than the value of the Type-2. Generally speaking, if the memory span $m$ of the NN model in Type-1 was made longer, the prediction accuracy could be improved. However, \Cref{tab:mse} shows that the estimation accuracy did not necessarily improve in large memory step $m$, which implied that using long distant past information might deteriorate the performance of the dynamic system. Considering the zigzag test. Zig-zag motion is a result of periodic control. Then, in its long periodicity might be considered. However, its long-distant memories were not generally essential for dynamical system modeling. Therefore, when predicting the maneuvering model, the memory span $m$ should be the span that is physically affected. In other words, if the memory span $m$ is an appropriate span that is not too long, the prediction accuracy will improve.
        
        \subsection{Comparison of Loss Function}\label{sec:LM}
            \begin{figure*}[t]
                \centering
                \includegraphics[width=2.0\columnwidth]{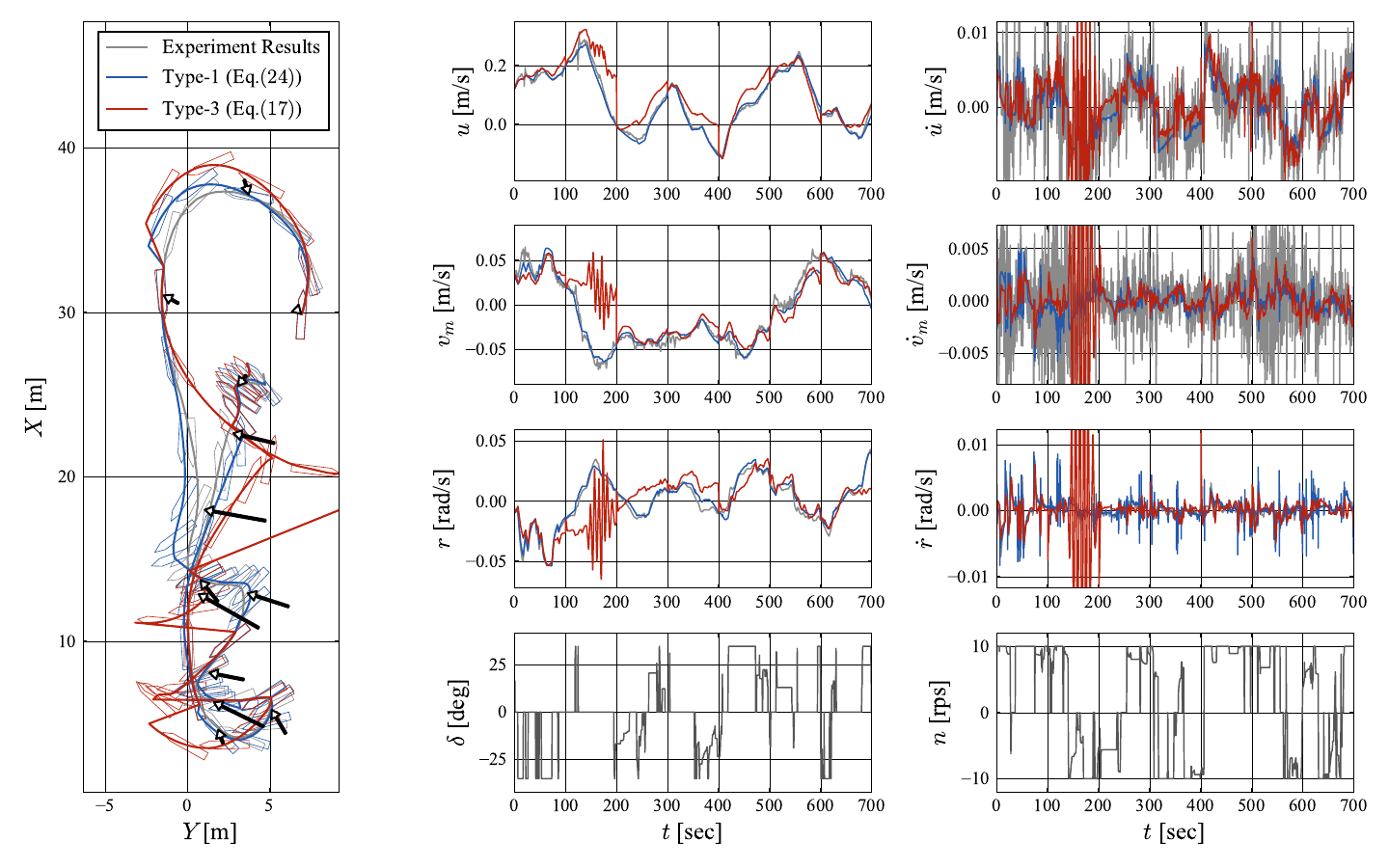}
                \caption{Comparison of the trajectories obtained for two loss functions [Eqs. (\ref{eq:stateloss}) and (\ref{eq:accloss})]. The training dataset is Train-TZRB+ in both methods.}
                \label{fig:LearningMethod}
            \end{figure*}

            In this section, comparing Type-1 and Type-3, the dependency of the prediction accuracy on the loss function was discussed. In Type-1, the proposed loss function [\cref{eq:stateloss}] was used. In Type-3, the loss function that directly evaluates acceleration [\cref{eq:accloss}] was used. \cref{eq:nnmodel2} was used as the NN model, and Train-TZRB+ was used as the data set. In addition to \Cref{tab:mse}, the simulation results of Type-1 and Type-3 in the random maneuver test dataset are shown in \Cref{fig:LearningMethod}.
            
            In \Cref{tab:mse}, Type-3 generally showed extreamly larger prediction error than  Type-1. This is because, in Random and Zigzag manuevers, the simulation of Type-3 shows the tendency of divergence.
            
            Besides, \Cref{fig:LearningMethod} shows that, in the proposed loss function, the predicted position and velocities almost coincided with those of the experimental results, whereas the accelerations did not show a good coincidence. Notably, as stated above, experimentally obtained accelerations, $\dot{u}$ and $\dot{v}_{m}$, were likely to include noises and spikes since these properties were derived from the time differentiation for the position data from GNSSs. Some parts of the spikes in experimental data were induced by data processing, i.e., time differentiation. These results showed that the proposed loss function [\Cref{eq:stateloss}] enabled us to predict the maneuvering model by avoiding the direct evaluation of accelerations.
            
            Moreover, the NN and experimental results of the yaw angular acceleration $\dot{r}$ coincided, and the tendency was not the same as the acceleration components, $\dot{u}$ and $\dot{v}_{m}$. In our data acquisition system, the yaw angular velocity was directly measured from FOG, and the yaw angular acceleration $\dot{r}$ was obtainable from one differential operation. Therefore, the spikes in the yaw angular acceleration $\dot{r}$ were likely to be lesser than those in the acceleration component. Therefore, the spikes that appeared in the experimental data of $\dot{r}$ were not necessarily noises.
            
            Therefore, we concluded that the proposed loss function [\Cref{eq:stateloss}] could reduce the effect of noise, thereby significantly improve the prediction accuracy of positions and velocities.
        
        \subsection{Comparison of Dataset}\label{sec:DS}
            In this section, comparing Type-1, Type-4, and Type-5, the dependency of prediction accuracy on the data set was assessed. In Type-1, Train-TZRB+, which has the largest data, was used as the dataset. In Type-4, Train-TZB, which did not contain random maneuvering test data, was used as the dataset. In Type-5, Train-TZRB, which contained the random maneuvering test data, was used as the dataset. To confirm the effectiveness of the random maneuvering test, Train-TZB and Train-TZRB have the same amount of data.
            
            From \Cref{tab:mse}, the performance of Type-1 was better than those of Type-4 and Type-5, which implied that a larger amount of dataset made the NN model more accurate. However, it is unrealistic to acquire an extremely large amount of training data. Thereofre, the balance of the amount of data and the prediction accuracy should be considered in practical applications.
            
            Comparing Type-4 and Type-5, in \Cref{tab:mse}, the prediction accuracy of the random maneuver test using the NN trained for Type-5, namely Train-TZRB, was better. Meanwhile, in the turning and zigzag tests, the prediction accuracy of the NN trained for Type-4, namely, Train-TZB, was better. These simply implied that the prediction accuracy for test datasets apparently depended on the inclusion of similar data in the training dataset. 
            
            From \Cref{tab:mse}, the result of Type-5 was better in Berthing1 and Berthing2 , and slightly better in Berthing3. In other words, in berthing test, which was the focus of this study, the prediction accuracy of Type-5 was generally better than that of Type-4. 
            Therefore, the inclusion of random maneuvering tests in the training dataset successfully made the prediction performance of NN better, even in the berthing control.
            
        \subsection{Comparison with MMG}\label{sec:M}
            
\begin{figure}[t]
                \centering
                \includegraphics[width=\columnwidth]{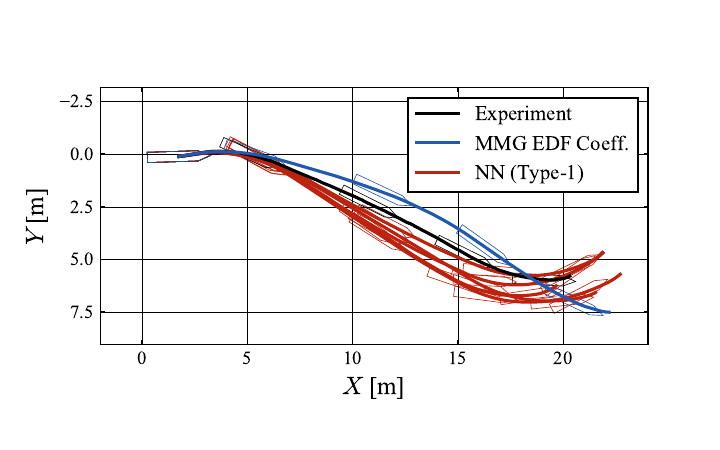}
                \caption{Comparison of prediction accuracy by each dataset in Berthing1 test. Time of test in model scale is $147.0 \mathrm{[s]}$}
                \label{fig:comparison_berthing1}
            \end{figure}

\begin{figure}[t]
                \centering
                \includegraphics[width=\columnwidth]{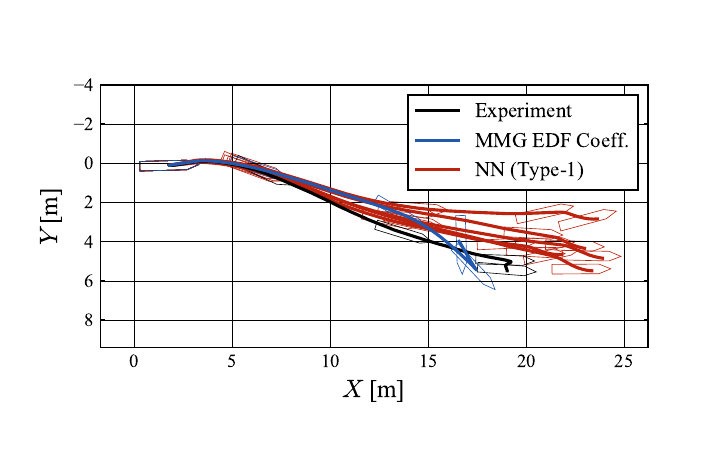}
                \caption{Comparison of prediction accuracy by each dataset in Berthing2 test. Time of test in model scale is $127.6 \mathrm{[s]}$}
                \label{fig:comparison_berthing2}
            \end{figure}

\begin{figure}[htbp]
                \centering
                \includegraphics[width=\columnwidth]{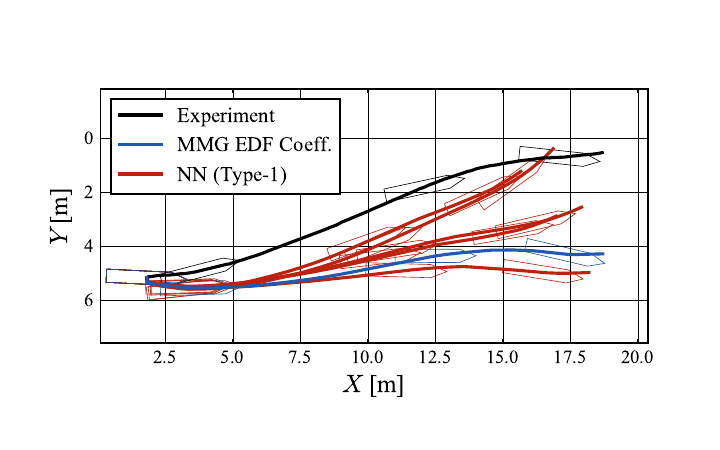}
                \caption{Comparison of prediction accuracy by each dataset in Berthing3 test. Time of test in model scale is $105.7 \mathrm{[s]}$}
                \label{fig:comparison_berthing3}
            \end{figure}

            In this section, the results using the MMG model and the NN model, particularly Type-1 and Type-5, were compared. The coefficients of the MMG model were determined by the captive model tests \cite{Miyauchi20201SI}. The simulation results of Type-1 and the MMG model in berthing test dataset are shown in \cref{fig:comparison_berthing1,fig:comparison_berthing2,fig:comparison_berthing3}. In Type-1, the results of five patterns with different random seeds was drawn.
            
            \Cref{tab:mse} shows that, in the random, turning, and zigzag tests, the prediction accuracies of Type-1 and Type-5 were better than that of the MMG model, which might be because the training dataset contained enough motion data similar to the test dataset. 
            
            Meanwhile, in Berthing1 and Berthing3, the prediction accuracy of the MMG model was better than those of Type-1 and Type-5, whereas the prediction accuracy of Type-1 and Type-5 was better than that of the MMG model in Berthing2, which could be considered that although maneuvers, which imitates the berthing motion were conducted in random maneuvering tests, the prepared training data failed to sufficiently cover the complex motion in actual berthing maneuver.
            
            However, the comparisons of all five trajectories for Type-1 are shown in \cref{fig:comparison_berthing1,fig:comparison_berthing2,fig:comparison_berthing3}. From these figures, although the results slightly depended on the random seeds, it was understood that the Type-1 results showed good prediction performance compared with the MMG model results. Moreover, increasing the percentage of berthing tests in the training data significantly contributes in improving the prediction accuracy of the berthing maneuvers.

        
    \section{Conclusion}
        
        In this research, the following five main outcomes were uncovered. 
        (i) Low-speed maneuvering model could be well approximated by RNN. 
        (ii) Comparison of two RNN models showed that the prediction accuracy of the RNN that ignores the memory before a certain time was better than the  ``standard'' RNN.
        (iii) To eliminate or reduce the adverse effect of the noise included in the measurement data, a loss function that evaluated possitions and velocities instead of acceleration directly was newly proposed. Then, the NN trained under the proposed loss function showed a significant improvement in prediction accuracy. 
        (iv) To obtain a better NN that can predict ship trajectories under berthing control, not only the usual maneuvering data, such as turning and zigzag tests, but also random maneuver tests were used in the training phase of NN. Then, it was confirmed that the inclusion of random maneuver tests in the training data significantly improved prediction performance.
        (v) The performance of the NN and MMG models were compared. Concerning random maneuvering, turning, and zigzag tests, which were sufficiently included in the training dataset, the prediction accuracy of the NN model was better than that of the MMG model. Meanwhile, concerning the berthing maneuvers, which were not sufficiently included in the training data, the prediction accuracy of the MMG model was better than the NN model.
        
        In the proposed prediction method using NN, the prediction accuracy strongly depended on the dataset, in particular the amount of data and types of manuevers in the data set. In addition, the prediction accuracy depended on random seeds. Therefore, the proposed method using RNN is not necessarily considered to be perfect. To make the proposed method more relizable and precise, preparation of the appropriate and sufficient amount of training data are essential. The detailed exploration for this point will be one of our future studies.

    \section*{Acknowledgements}
        This paper is a preprint published in the Journal of Marine Science and Technology, and the public version is available at (https://link.springer.com/article/10.1007/s00773-021-00867-1). This study was supported by a Grant-in-Aid for Scientific Research from the Japan Society for Promotion of Science (JSPS KAKENHI Grant \#19K04858). The study also received assistance from JFY2018 Fundamental Research Developing Association for Shipbuilding and Offshore (REDAS) in Japan. Authors also would like to express gratitude to Mr. Satoru Konishi, Magellan Systems Japan Inc., to the technical support on GNSS measurement during the free run model test. 


\begin{thebibliography}{10}
    
    \bibitem{whitepaper2020}
    {Ministry of Land, Infrastructure, Transport and Tourism}.
    \newblock {White paper on land, infrastructure, transport and tourism in
      Japan},
    \newblock 2020.
    
    \bibitem{Kose1984}
    Kuniji Kose, Hiroyoshi Hinata, Yasuhisa Hashizume, and Eijiro Futagawa.
    \newblock On a mathematical model of maneuvering motions of ships in low
      speeds.
    \newblock {\em Journal of the Society of Naval Architects of Japan},
      155:132--138, 1984.
    
    \bibitem{yamato1990}
    H. Yamato.
    \newblock Automatic berthing by the neural controller.
    \newblock {\em Proceedings of Nineth ship control systems Symposium}, 3:3183--3201,
      1990.
    
    \bibitem{Shouji1992}
    Kouichi Shouji and Kohei Ohtsu.
    \newblock A study on the optimization of ship maneuvering by optimal control
      theory (1st report).
    \newblock {\em Journal of the Society of Naval Architects of Japan},
      172:365--373, 1992.
    
    \bibitem{hasegawa1993}
    Kazuhiko Hasegawa and Keiji Kitera.
    \newblock Automatic berthing control system using network and knowledge-base.
    \newblock {\em Journal of the Kansai Society of Naval Architects, Japan},
      220:135--143, 1993.
    
    \bibitem{Maki2020}
    Atsuo Maki, Naoki Sakamoto, Youhei Akimoto, Hiroyuki Nishikawa, and Naoya
      Umeda.
    \newblock {Application of optimal control theory based on the evolution
      strategy (CMA-ES) to automatic berthing}.
    \newblock {\em Journal of Marine Science and Technology}, 25(1):221--233, 2020.
    
    \bibitem{Miyauchi2021}
    Yoshiki Miyauchi, Ryohei Sawada, Youhei Akimoto, Naoya Umeda, and Atsuo Maki.
    \newblock Optimization on planning of trajectory and control of autonomous
      berthing and unberthing for the realistic port geometry.
    \newblock {\em Ocean Engineering}.
    
    \bibitem{Dimas2020}
    Dimas~M. Rachman, A. Maki, and N. Umeda.
    \newblock Numerical simulation of automatic berthing by cma-es in real time.
    \newblock {\em Conference proceedings, the Japan Society of Naval Architects
      and Ocean Engineers (in Japanese)}, 11 2020.
    
    \bibitem{Dimas2021}
    Dimas~M. Rachman, Naoya Umeda, Atsuo Maki, and Miyauchi Yoshiki.
    \newblock Feasibility study on the use of evolution strategy: Cma-es for ship
      automatic docking problem.
    \newblock 2021.
    
    \bibitem{Akimoto2021}
    Youhei Akimoto, Yoshiki Miyauchi, and Atsuo Maki.
    \newblock Saddle point optimization with approximate minimization oracle and
      its application to robust berthing control, 2021.
    
    \bibitem{wakita2021}
    K.~Wakita, Y.~Akimoto, K.~Shoji, Y.~Miyauchi, N.~Umeda, and A.~Maki.
    \newblock On transfer learning of the pre-trained policy of ship tracking
      control in a real environment.
    \newblock {\em Conference proceedings, Japan Society of Naval Architects
      and Ocean Engineers (in Japanese)}, May 2021.
    
    \bibitem{Abkowitz1980}
    M.~Abkowitz.
    \newblock Measurement of hydrodynamic characteristics from ship maneuvering
      trials by system identification.
    \newblock 1980.
    
    \bibitem{yasukawa2015}
    H.~Yasukawa and Y.~Yoshimura.
    \newblock Introduction of mmg standard method for ship maneuvering predictions.
    \newblock {\em Journal of Marine Science and Technology}, 20(1):37--52, 2015.
    
    \bibitem{SAKAMOTO2019}
    Nobuaki Sakamoto, Kunihide Ohashi, Motoki Araki, Kenichi Kume, and Hiroshi
      Kobayashi.
    \newblock Identification of kvlcc2 manoeuvring parameters for a modular-type
      mathematical model by rans method with an overset approach.
    \newblock {\em Ocean Engineering}, 188:106257, 2019.
    
    \bibitem{SUTULO2014}
    Serge Sutulo and C.~{Guedes Soares}.
    \newblock An algorithm for offline identification of ship manoeuvring
      mathematical models from free-running tests.
    \newblock {\em Ocean Engineering}, 79:10 -- 25, 2014.
    
    \bibitem{ARAKI2012}
    Motoki Araki, Hamid Sadat-Hosseini, Yugo Sanada, Kenji Tanimoto, Naoya Umeda,
      and Frederick Stern.
    \newblock Estimating maneuvering coefficients using system identification
      methods with experimental, system-based, and cfd free-running trial data.
    \newblock {\em Ocean Engineering}, 51:63 -- 84, 2012.
    
    \bibitem{Miyauchi2020}
    Y.~Miyauchi, A.~Maki, N.~Umeda, M. R. Dimas, T.~Shimoji, K.~Wakita, and
      Y.~Akimoto.
    \newblock On system identification for low-speed maneuvering model by using
      cma-es (4th report).
    \newblock {\em Conference proceedings, Japan Society of Naval Architects
      and Ocean Engineers (in Japanese)}, 2020.
    
    \bibitem{MUNOZMANSILLA2009}
    Roc\'{i}o~Mu\ {n}oz Mansilla, Joaqu\'{i}n Aranda, Jos\'{e}~Manuel D\'{i}az, and
      J~{de la Cruz}.
    \newblock Parametric model identification of high-speed craft dynamics.
    \newblock {\em Ocean Engineering}, 36(12):1025--1038, 2009.
    
    \bibitem{Nagumo1967}
    J.~{Nagumo} and A.~{Noda}.
    \newblock A learning method for system identification.
    \newblock {\em IEEE Transactions on Automatic Control}, 12(3):282--287, June
      1967.
    
    \bibitem{KALLSTROM1981}
    C.G. K\"{a}llstr\"{o}m and K.J.~\AA str\"{o}m.
    \newblock Experiences of system identification applied to ship steering.
    \newblock {\em Automatica}, 17(1):187--198, 1981.
    
    \bibitem{ASTROM1980}
    K.J.~\AA str\"{o}m.
    \newblock Maximum likelihood and prediction error methods.
    \newblock {\em Automatica}, 16(5):551--574, 1980.
    
    \bibitem{Perera2015}
    Lokukaluge~P. Perera, P.~Oliveira, and C.~{Guedes Soares}.
    \newblock {System identification of nonlinear vessel steering}.
    \newblock {\em Journal of Offshore Mechanics and Arctic Engineering}, 137(3),
      jun 2015.
    
    \bibitem{Yasukawa2021}
    H.~Yasukawa, T.~Ishikawa, and Y.~Yoshimura.
    \newblock {Investigation on the rudder force of a ship in large drifting
      conditions with the MMG model}.
    \newblock {\em Journal of Marine Science and Technology (Japan)}, 2021.
    
    \bibitem{XU2020}
    Haitong Xu, M.A. Hinostroza, Zihao Wang, and C.~{Guedes Soares}.
    \newblock Experimental investigation of shallow water effect on vessel steering
      model using system identification method.
    \newblock {\em Ocean Engineering}, 199:106940, 2020.
    
    \bibitem{LUO2014}
    Weilin Luo, Lúcia Moreira, and C.~{Guedes Soares}.
    \newblock Manoeuvring simulation of catamaran by using implicit models based on
      support vector machines.
    \newblock {\em Ocean Engineering}, 82:150--159, 2014.
    
    \bibitem{Luo2013SVM}
    {\em {Parameter Identification of Ship Manoeuvring Model Based on Particle
      Swarm Optimization and Support Vector Machines}}, volume 5: Ocean
      Engineering of {\em International Conference on. Journal of Offshore Mechanics and Arctic
      Engineering}, 06, 2013.
    \newblock V005T06A071.
    
    \bibitem{ZHU2020}
    Man Zhu, Wuqiang Sun, Axel Hahn, Yuanqiao Wen, Changshi Xiao, and Wei Tao.
    \newblock Adaptive modeling of maritime autonomous surface ships with
      uncertainty using a weighted ls-svr robust to outliers.
    \newblock {\em Ocean Engineering}, 200:107053, 2020.
    
    \bibitem{Mei2019}
    B.~{Mei}, L.~{Sun}, and G.~{Shi}.
    \newblock White-black-box hybrid model identification based on rm-rf for ship
      maneuvering.
    \newblock {\em IEEE Access}, 7:57691--57705, 2019.
    
    \bibitem{MOREIRA2003}
    L.~Moreira and C.~{Guedes Soares}.
    \newblock Dynamic model of manoeuvrability using recursive neural networks.
    \newblock {\em Ocean Engineering}, 30(13):1669 -- 1697, 2003.
    
    \bibitem{OSKIN2013}
    Dmitry~A. Oskin, Alexander~A. Dyda, and Vasily~E. Markin.
    \newblock Neural network identification of marine ship dynamics.
    \newblock {\em IFAC Proceedings Volumes}, 46(33):191 -- 196, 2013.
    \newblock 9th IFAC Conference on Control Applications in Marine Systems.
    
    \bibitem{Chiu2004}
    {Forng-Chen Chiu}, {Tun-Li Chang}, {Jenhwa Go}, {Shean-Kwang Chou}, and
      {Wei-Chung Chen}.
    \newblock A recursive neural networks model for ship maneuverability
      prediction.
    \newblock In {\em Oceans MTS/IEEE Techno-Ocean '04 (IEEE Cat.
      No.04CH37600)}, volume~3, pages 1211--1218 Vol.3, Nov 2004, '04.
    
    \bibitem{RAJESH2008}
    G.~Rajesh and S.K. Bhattacharyya.
    \newblock System identification for nonlinear maneuvering of large tankers
      using artificial neural network.
    \newblock {\em Applied Ocean Research}, 30(4):256--263, 2008.
    
    \bibitem{Luo2016NN}
    Weilin Luo and Zhicheng Zhang.
    \newblock {Modeling of ship maneuvering motion using neural networks}.
    \newblock {\em Journal of Marine Science and Application}, 15(4):426--432,
      2016.
    
    \bibitem{Koda2020-1}
    T.~Koda, K.~Furutachi, Y.~Furukawa, and H.~Ibaragi.
    \newblock Development of rnn-based prediction model for ship manoeuvring
      motion.
    \newblock {\em Conference proceedings, 30, Japan Society of Naval Architects
      and Ocean Engineers}, pages 609--612, May 2020.
    
    \bibitem{Koda2020-2}
    T.~Koda, K.~Furutachi, Y.~Furukawa, and H.~Ibaragi.
    \newblock Development of rnn-based prediction model for ship manoeuvring motion
      under external disturbances.
    \newblock {\em Conference proceedings, 31, Japan Society of Naval Architects
      and Ocean Engineers}, 2020.
    
    \bibitem{FUNAHASHI1989183}
    Ken-Ichi Funahashi.
    \newblock On the approximate realization of continuous mappings by neural
      networks.
    \newblock {\em Neural Networks}, 2(3):183 -- 192, 1989.
    
    \bibitem{Cybenko1989}
    G.~Cybenko.
    \newblock {Approximation by superpositions of a sigmoidal function}.
    \newblock {\em Mathematics of Control, Signals, and Systems}, 2(4):303--314,
      1989.
    
    \bibitem{HORNIK1991251}
    Kurt Hornik.
    \newblock Approximation capabilities of multilayer feedforward networks.
    \newblock {\em Neural Networks}, 4(2):251 -- 257, 1991.
    
    \bibitem{rumelhart1985}
    David~E. Rumelhart, Geoffrey~E. Hinton, and Ronald~J. Williams.
    \newblock Learning internal representations by error propagation.
    \newblock Technical report, California Univ San Diego La Jolla Inst for
      Cognitive Science, 1985.
    
    \bibitem{rosenblatt1958}
    Frank Rosenblatt.
    \newblock The perceptron: A probabilistic model for information storage and
      organization in the brain.
    \newblock {\em Psychological Review}, 65(6):386-408, 1958.
    
    \bibitem{Krizhevsky2012}
    Alex Krizhevsky, Ilya Sutskever, and E.~Geoffrey. Hinton.
    \newblock Imagenet classification with deep convolutional neural networks.
    \newblock In Advances in neural information processing systems, 25 F.~Pereira, C.~J.~C. Burges, L.~Bot-Tou, and K.~Q. Weinberger,
      editors, Curran Associates, Inc, pages 1097--1105, 2012.
    
    \bibitem{Vinyals2015}
    Oriol Vinyals and Quoc~V. Le.
    \newblock A neural conversational model.
    \newblock {\em Clinical Orthopaedics and Related Research}, Abs/1506.05869, 2015.
    
    \bibitem{Raghavendra2018}
    Raghavendra Chalapathy, Aditya~Krishna Menon, and Sanjay Chawla.
    \newblock Anomaly detection using one-class neural networks.
    \newblock {\em Clinical Orthopaedics and Related Research}, Abs/1802.06360, 2018.
    
    \bibitem{Nakanishi1997}
    Hiroaki Nakanishi, Takehisa Kohda, and Koichi Inoue.
    \newblock A design method of optimal state feed-back control systems by use of
      neural network.
    \newblock {\em Transactions of the Society of Instrument and Control
      Engineers}, 33(9):882--889, 1997.
    
    \bibitem{kingma2014}
    Diederik~P. Kingma and Jimmy Ba.
    \newblock Adam: A method for stochastic optimization.
    \newblock {\em arXiv preprint arXiv:1412.6980}, 2014.
    
    \bibitem{Yoshimura2009}
    Yasuo Yoshimura, Ikao Nakao, and Atsushi Ishibashi.
    \newblock Unified mathematical model for ocean and harbour manoeuvring.
    \newblock International Conference on Marine Simulation and
      Ship Maneuverability, pages 116--124, August 2009.
    
    \bibitem{Yasukawa2008}
    Hironori Yasukawa.
    \newblock {Simulations of ship maneuvering in waves}.
    \newblock {\em Journal of the Japan Society of Naval Architects and Ocean
      Engineers}, 7(0):163--170, 2008.
    
    \bibitem{Hachii2004}
    Tomoyuki Hachii.
    \newblock The prediction of manoeuvring motion on ships with low speed using
      standard mmg model.
    \newblock Master's thesis, Osaka University (in Japanese), 2004.
    
    \bibitem{Kitagawa2015}
    Yasushi Kitagawa, Yoshiaki Tsukada, and Hideki Miyazaki.
    \newblock {GS1-1 A study on mathematical models of propeller and rudder
      under maneuvering with propeller reverse rotation}.
    \newblock {\em Conference Proceedings The Japan Society of Naval Architects and
      Ocean Engineers}, 20, pages 117--120, 2015s.
    
    \bibitem{Miyauchi20201SI}
    Yoshiki Miyauchi, Atsuo Maki, Naoya Umeda, Dimas~M. Rachman, and Youhei
      Akimoto.
    \newblock System parameter exploration of ship maneuvering model for automatic
      docking / berthing using cma-es.
    \newblock {\em Journal of Marine Science and Technology (to be submitted)},
      2021.
    
    \bibitem{Wada2019}
    Suisei Wada, Naoya Umeda, and Atsuo Maki.
    \newblock Development of general purpose free-running model ship with ros :
      Enhanced model ship experiments system.
    \newblock {\em Conference proceedings, the Japan Society of Naval Architects
      and Ocean Engineers (in Japanese)}, (28):587--594, Jun 2019.
    
    \end{thebibliography}

\end{document}